\documentclass[12pt]{article}
\usepackage{graphicx}
\usepackage{centernot}

\usepackage{mathtools}
\DeclarePairedDelimiter\bra{\langle}{\rvert}
\DeclarePairedDelimiter\ket{\lvert}{\rangle}
\DeclarePairedDelimiterX\braket[2]{\langle}{\rangle}{#1 \delimsize\vert #2}

\def\ignore#1{{}}

\let\oldtheequation=\theequation
\def\doteqs#1{\setcounter{equation}{0}            
\def\theequation{{#1}.\oldtheequation}}
\newcounter{sxn}
\def\sx#1{\addtocounter{sxn}{1} \vskip 1.cm  \goodbreak
\noindent{\large\bf\leftline{\thesxn.~~#1}} \nobreak \vskip -.5cm}
\def\sxn#1{\sx{#1} \doteqs{\thesxn}}

\newcounter{axn}

\date{}

\newdimen\mybaselineskip
\newcommand{\beeq}{\begin{equation}}
\newcommand{\eneq}{\end{equation}}
\newcommand{\beqn}{\begin{eqnarray}}
\newcommand{\eeqn}{\end{eqnarray}}

\def\la{\raise.16ex\hbox{$\langle$}\lower.16ex\hbox{}  }
\def\ra{\, \raise.16ex\hbox{$\rangle$}\lower.16ex\hbox{} }

\def\psibar{ \psi \kern-.65em\raise.6em\hbox{$-$} \lower.6em\hbox{} }
\def\psibarb{ \psi \kern-.65em\raise.6em\hbox{$-$}  }

\begin{document}

\thispagestyle{empty}

\baselineskip=12pt

\vspace*{3.cm}

\begin{center}  
{\LARGE \bf  A Modified Stern-Gerlach Experiment Using a Quantum Two-State Magnetic Field}
\end{center}

\baselineskip=14pt

\vspace{3cm}
\begin{center}
{\bf  Ramin G.~Daghigh$\sharp$,  Michael D.~Green$\dagger$ and Christopher J.~West$\ddagger$}
\end{center}

\centerline{\small \it $\sharp$ Natural Sciences Department, Metropolitan State University, Saint Paul, Minnesota, USA 55106}
\vskip 0 cm
\centerline{}

\centerline{\small \it $\dagger$ Mathematics Department, Metropolitan State University, Saint Paul, Minnesota, USA 55106}
\vskip 0 cm
\centerline{}

\centerline{\small \it $\ddagger$ Center for Academic Excellence, Metropolitan State University, Saint Paul, Minnesota, USA 55106}
\vskip 0 cm
\centerline{} 

\centerline{\small \it $\ddagger$ Joint Institute for Nuclear Astrophysics, University of Notre Dame, Notre Dame, Indiana, USA 46545}
\vskip 0 cm
\centerline{} 

\vspace{1cm}
\begin{abstract}

The Stern-Gerlach experiment has played an important role in our understanding of quantum behavior. We propose and analyze a modified version of this experiment where the magnetic field of the detector is in a quantum superposition, which may be experimentally realized using a superconducting flux qubit. We show that if incident spin-$1/2$ particles couple with the two-state magnetic field, a discrete target distribution results that resembles the distribution in the classical Stern-Gerlach experiment. As an application of the general result, we compute the distribution for a square waveform of the incident fermion. This experimental setup allows us to establish: (1) the quantization of the intrinsic angular momentum of a spin-$1/2$ particle, and (2) a correlation between EPR pairs leading to nonlocality, without necessarily collapsing the particle's spin wavefunction.  

\baselineskip=20pt plus 1pt minus 1pt
\end{abstract}

\newpage

\sxn{Introduction}
\vskip 0.3cm

Quantum mechanics has been used to study and predict the behavior of quantum systems using classical detectors that do not display quantum or nonlocal behavior. In doing so, nonlocal systems are studied using detectors that strictly follow the principle of locality.  It seems natural to ask what happens if we explore the measurements of quantum particles/systems using detectors that also display quantum behavior.  

Such investigations have precedence. For example, the use of quantum control devices has been proposed and implemented in a series of experiments designed to determine if Bohr's complementarity principle requires modification \cite{ionicioiu2011,auccaise2012,roy2012,peruzzo2012,kaiser2012,tang2012,tang2013,qureshi2008}; see also the review article \cite{xiao2016} (and references therein) for the specific application to delayed choice experiments. In the proposed experimental setup \cite{ionicioiu2011}, for example, the quantum control device is a Mach-Zehnder interferometer with a beam splitter in a superposition of present or absent states, thereby intending to observe the wave and particle behavior of incident photons simultaneously. 

In the present work, we propose a modification of the Stern-Gerlach experiment where the magnetic field of the detector is in a superposition of two quantum states.  This can be experimentally realized, for example, by using a micro-meter sized superconducting loop with Josephson-junctions (such as the one described in \cite{Qubit1, Qubit2}) that is used as the core component of superconducting flux qubits. The simplicity of this setup makes it a good candidate to investigate both the fundamental aspects of quantum measurement and the consequences of using detectors with quantum behavior. The present work explores the setup in a gedanken experiment, but the setup is experimentally viable considering the progress made in the study of qubits. 

The use of a superconducting flux qubit in a Stern-Gerlach experiment was originally proposed in the first version of this paper\cite{firstdraft}.  Shortly afterward, Singh\cite{Singh} explored a similar setup that uses a superconducting flux qubit to create a quantum entanglement involving macroscopic components. In that work, the incident ``particle'' is a Bose-Einstein condensate \textit{drip}. Through entanglement with the loop's magnetic field, the condensate remains in a superposition of spatial coordinates. In \cite{Singh}, however, the incident Bose-Einstein condensate drip has only one polarization. Thus, the setup bears similarity to the classical Stern-Gerlach experiment in the reference frame of the condensate.  In the present work, however, both the incident particle and the superconducting loop exist in mixed states and entangle to form triplet and singlet spin states.  Thus, it is not known \textit{a priori} what resulting distribution will emerge given the results of \cite{Singh}. We compute the outcome of the entangled particle-loop states using a perturbative approach.  

In the following Section, we explain the setup and then detail how the magnetic field is constructed. In Section 3, we derive a general form for the discrete target distribution from the Hamiltonian. In Section 4, we briefly discuss our numerical calculations for a chosen incident waveform. Finally, we offer a conclusion exploring implications of our results.  

\sxn{The Modified Stern-Gerlach Apparatus}
\vskip 0.3cm

We suggest an apparatus that consists of a superconducting loop interrupted by Josephson-junctions, e.g., \cite{Qubit1, Qubit2}. When subjected to a small magnetic field, persistent clockwise or counterclockwise currents are induced. The loop behaves as a particle in a double-well potential with two classical states comprised of clockwise or counterclockwise persistent-currents located at the potential's minima. This double-well potential is symmetric when the enclosed quantized flux has half-integer multiples of $\Phi_0=h/(2e)$, resulting in both classical states being stable. In this configuration, we achieve a superposition of persistent clockwise and counterclockwise currents, yielding a magnetic dipole in a superposition of two opposite directions. For other values of the enclosed flux, the double-well potential is non-symmetric with only one stable classical state.  The superposition of macroscopic persistent-current states in the superconducting loop proposed in \cite{Qubit1, Qubit2} is realized experimentally, for example, in \cite{Qubit3} with the measured persistent-current of $450 \pm 50\,$nA. Additional discussion/review of flux qubits can be found in \cite{devoret2004a,wendin2007,QubitReview2,clarke2008}.  Qubit use is also extensive in quantum computing; see \cite{devoret2004b,you2005,wendin2005,ladd2010} for more information on this topic.

\begin{figure}[h]
	\begin{center}
		\includegraphics[height=8cm]{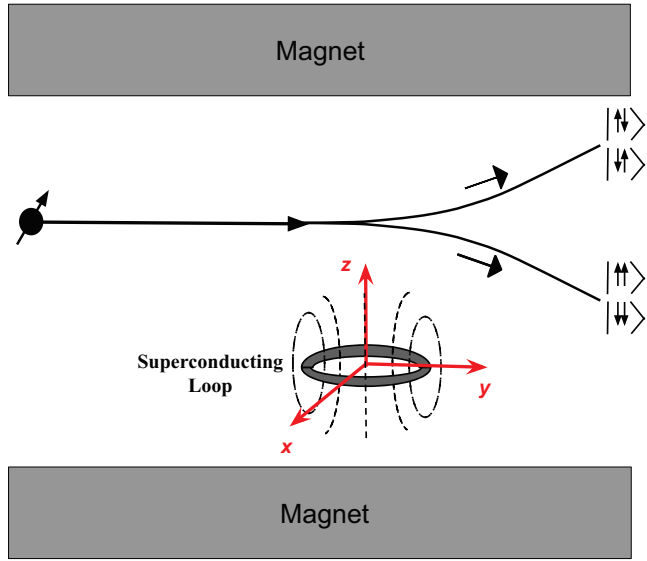}
	\end{center}
	\caption{Schematic presentation of proposed modified Stern-Gerlach experiment. A superconducting loop is placed in a uniform magnetic field to produce a superposition of magnetic dipole directions. The dipole couples with an incident fermion to produce a discrete distribution on a background screen. The combined spin state is denoted by $\left| \mbox{particle, loop}\right\rangle$. Experimental implementation should account for possible edge effects from the magnets.}
	\label{QubitDetector}
\end{figure}

The magnetic field in our modified Stern-Gerlach apparatus can be produced by a superconducting loop as described above. The loop is placed in a \textit{uniform} external magnetic field ${\bf B}_0$, tuned to produce a flux through the superconducting loop with half-integer multiples of $\Phi_0$. The uniform magnetic field direction also defines the $z$-axis for the experimental setup. This setup produces a magnetic field in a superposition of two quantum states. Analogous to the classical Stern-Gerlach experiment, an incident spin-$1/2$ particle is deflected by the magnetic field of the loop and hits a screen on the other side (Figure \ref{QubitDetector}). Note the external magnetic field is uniform and hence does not deflect the incoming particle. 

This superconducting loop is a macroscopic system with quantum behavior. In the following discussion, however, we do not demonstrate whether interaction with the incident fermion will in fact lead to entangled spin states rather than a classical measurement (i.e., the particle's spin wavefunction collapsing, or the loop's, or both). Rather, we assume the established quantum laws are applicable to macroscopic quantum systems.  See \cite{Singh} for details/discussions on the necessary conditions required for entanglement. In addition, a review of qubit interaction with other quantum systems can be found in \cite{xiang2013} (and references therein). See also \cite{shnyrkov2007} for a comparison of ScS- and SIS-based qubit systems used as quantum detectors. 

Note that whereas a superconducting loop is a good candidate for this experiment, the following analysis is generalizable to any system in a superposition of two magnetic field directions.  

\sxn{Analysis and Discussion}
\vskip 0.3cm

First, we build the Hamiltonian for the particle-loop system. Next, an expression for the expectation values of the particle's position along the $z$-axis is computed using a perturbation expansion of the solution to Schr$\ddot{\mbox{o}}$dinger's equation. Our goal is to determine the qualitative nature of the distribution pattern of the particles; achieving an exact solution to this problem is beyond the scope of this work\footnote{Even though the Stern-Gerlach experiment was conducted in 1921-22, an approximation-free solution was not found until 2011 by \cite{Hsu}.}. The perturbative method also allows for an order of magnitude estimation of the particle's deflection to gauge experimental viability.  See the next section for details. 

The magnetic dipole moment generated by the superconducting loop or any other two-state magnetic system can be written as

\beeq
{\bf{m}}^{(l)}= \beta {\bf S}^{(l)}~,  
\label{ml}
\eneq

\noindent where ${\bf S}^{(l)}$ is the spin vector and  $\beta$ is a constant of proportionality.  Similarly, the magnetic dipole moment of the incident spin-$1/2$ particle to be detected is described as

\beeq
{\bf m}^{(p)}= \alpha {\bf S}^{(p)}~,
\label{mp}
\eneq

\noindent where ${\bf S}^{(p)}$ is the spin of the particle and $\alpha$ is a constant of proportionality.  The magnetic field created by the loop can be stated as

\beeq
{\bf B}^{(l)} = \frac{\mu_0}{4\pi r^3}\left[\frac{3}{r^2}(\textbf{m}^{(l)} \cdot  {\bf{r}}){\bf{r}}-\textbf{m}^{(l)}\right]+\frac{2\mu_0}{3} \textbf{m}^{(l)} \delta^3(\textbf{r})~,
\label{b-field}
\eneq

\noindent where $\mu_0$ is the vacuum permeability. The interaction Hamiltonian between the incident particle and the mixed field of the loop is given by the familiar relation

\begin{eqnarray}
H_{int.} &=& - {\bf {m}}^{(p)} \cdot \textbf{B}^{(l)}  \nonumber \\
&=&  - \frac{\mu_0}{4\pi r^3}\left[\frac{3}{r^2}(\textbf{m}^{(p)} \cdot  {\bf {r}})({\bf {m}}^{(l)} \cdot{\bf{r}})-{\bf {m}}^{(p)} \cdot \textbf{m}^{(l)}\right] \nonumber \\
&& - \frac{2\mu_0}{3} \textbf{m}^{(p)} \cdot \textbf{m}^{(l)} \delta^3(\textbf{r})~.
\label{int-H}
\end{eqnarray}

The complete Hamiltonian includes the kinetic and external uniform magnetic field terms and replaces the magnetic dipole moments with spin operators according to (\ref{ml}) and (\ref{mp}),
\begin{eqnarray}
\hat{H} &=& I \frac{\hat{p}^2}{2m}-\left(\alpha \hat{S}^{(p)}_{z} B_0 +\beta \hat{S}^{(l)}_{z} B_0\right) - \frac{\mu_0 \alpha \beta}{4\pi \hat{r}^3}\left[\frac{3}{\hat{r}^2} (\hat{{\bf S}}^{(p)}\cdot \hat{\bf{r}})(\hat{{\bf S}}^{(l)}\cdot \hat{\bf{r}})-\hat{{\bf S}}^{(p)} \cdot \hat{{\bf S}}^{(l)}\right] \nonumber \\
&&  -\frac{2\mu_0}{3} \alpha \beta \hat{\textbf{S}}^{(p)} \cdot \hat{\textbf{S}}^{(l)} \delta^3(\hat{\textbf{r}})~,
\label{total-H}
\end{eqnarray}

\noindent where $\hat{p}$ is the momentum operator for the particle with respect to the loop, $\hat{r}$ is the position operator, $I$ is the $2 \times 2$ identity matrix, operators are identified using carets, and vectors are denoted in bold. Note that we are performing the calculations in the loop frame, which may be stationary in the lab frame. In general, however, all stated position operators provide measurements for the particle-loop system in relative coordinates. The Schr$\ddot{\mbox{o}}$dinger equation to be solved is

\beeq
\hat{H} \Psi = i \hbar \frac{\partial \Psi}{\partial t}~,
\label{Scrodinger}
\eneq

\noindent where the loop-particle wavefunction $\Psi=\Psi(x,y,z,t)$ involves the two-component spinors

\beeq
\Psi = \left (\begin{array}{ll}  \psi^p_\uparrow \\ \\ \psi^p_\downarrow \end{array} \right) \otimes \left (  \begin{array}{ll}  \psi^l_\uparrow \\ \\ \psi^l_\downarrow \end{array} \right )~,
\label{Psi}
\eneq

\noindent with the spinor components given by

\begin{eqnarray}
&& \Psi_{\uparrow\uparrow} = \left(\begin{array}{ll} \psi^p_\uparrow \\ 0 \end{array} \right) \otimes \left (\begin{array}{ll} \psi^l_\uparrow \\ 0 \end{array} \right),~
\Psi_{\uparrow\downarrow} = \left(\begin{array}{ll} \psi^p_\uparrow \\ 0 \end{array} \right) \otimes \left (\begin{array}{ll} 0 \\ \psi^l_\downarrow \end{array} \right), \nonumber \\
&& \Psi_{\downarrow\uparrow} = \left(\begin{array}{ll} 0 \\ \psi^p_\downarrow \end{array} \right) \otimes \left (\begin{array}{ll} \psi^l_\uparrow \\ 0 \end{array} \right),~
\Psi_{\downarrow\downarrow} = \left(\begin{array}{ll} 0 \\ \psi^p_\uparrow \end{array} \right) \otimes \left (\begin{array}{ll} 0 \\ \psi^l_\downarrow \end{array} \right)~.
\label{Psi-4}
\end{eqnarray}

\noindent The deflection of incoming particles in the $z$-direction is given by the expectation values along the $z$-axis,

\begin{eqnarray}
{\left\langle \hat{z}(t)\right\rangle}_{\uparrow\uparrow}  &=& \int \Psi^\dagger_{\uparrow\uparrow}(t) \hat{z}  \Psi_{\uparrow\uparrow}(t) d^3r   \nonumber \\
{\left\langle \hat{z}(t)\right\rangle}_{\uparrow\downarrow}  &=& \int \Psi^\dagger_{\uparrow\downarrow}(t) \hat{z}  \Psi_{\uparrow\downarrow}(t) d^3r   \nonumber \\
{\left\langle \hat{z}(t)\right\rangle}_{\downarrow\uparrow}  &=& \int \Psi^\dagger_{\downarrow\uparrow}(t) \hat{z}  \Psi_{\downarrow\uparrow}(t) d^3r   \nonumber \\
{\left\langle \hat{z}(t)\right\rangle}_{\downarrow\downarrow}  &=& \int \Psi^\dagger_{\downarrow\downarrow}(t) \hat{z}  \Psi_{\downarrow\downarrow}(t) d^3r ~,
\label{z-expect}
\end{eqnarray}

\noindent where the Hamiltonian enters as the propagator on the stationary states,

\beeq
\Psi (t) = e^{-i \hat{H} t/\hbar} \Psi(0)~.
\eneq

The object $ e^{+i \hat{H} t/\hbar}\hat{z}e^{-i \hat{H} t/\hbar}$ can be evaluated perturbatively using the method suggested in \cite{Bulnes} where it is applied to the classical Stern-Gerlach experiment\footnote{In \cite{Bulnes} the authors use a ``reduced" Hamiltonian, but we do not.}. In general, the perturbative expansion is presented using commutators in the Baker-Campbell-Hausdorff formula,

\beeq
e^{\hat{O}} \hat{z} e^{-\hat{O}}= \hat{z} + [\hat{O}, \hat{z}] + \frac{1}{2!}[\hat{O}, [\hat{O}, \hat{z}]]+ \frac{1}{3!}[\hat{O},[\hat{O}, [\hat{O}, \hat{z}]]]+ \cdots~,
\label{Baker-H}
\eneq

\noindent where for our purposes

\beeq
\hat{O} = i \hat{H} t/\hbar~.
\label{bigO}
\eneq

\noindent The expectation values then take the form

\beeq
\left\langle\hat{z}(t)\right\rangle = \int \Psi^\dagger(0)\hat{z}\Psi(0) d^3r + \int \Psi^\dagger(0)[\hat{O}, \hat{z}]\Psi(0) d^3r + \int \Psi^\dagger(0)\frac{1}{2!}[\hat{O}, [\hat{O}, \hat{z}]]\Psi(0) d^3r+\cdots~,
\label{zt}
\eneq

\noindent whereby we may use Ehrenfest's theorem in the present context:

\beeq
{\left\langle \hat{z}(t)\right\rangle}  =  {\left\langle \hat{z}(0)\right\rangle}  + {\left\langle \hat{v}_z(0)\right\rangle} t +\frac{1}{2} \left\langle \hat{a}_z\right\rangle t^2+\cdots~,
\label{constant-a}
\eneq

\noindent to identify the commutator terms

\begin{eqnarray}
\left\langle \hat{v}_z(0)\right\rangle = \frac{\left\langle[\hat{O}, \hat{z}]\right\rangle}{t},~ \\
\left\langle \hat{a}_z\right\rangle = \frac{\left\langle [\hat{O}, [\hat{O},\hat{z}]]\right\rangle}{t^2}.
\end{eqnarray}

Computing the deflection thus involves evaluating the expectation value of the commutator recognized as the ``force" term, $\left\langle [\hat{O}, [\hat{O},\hat{z}]]\right\rangle m/t^2$. We proceed by evaluating the commutators in (\ref{Baker-H}):
\beeq
[\hat{O}, \hat{z}] =  \frac{\hat{p}_z}{m} t~,
\label{commute-1}
\eneq
\begin{eqnarray}
[\hat{O}, [\hat{O}, \hat{z}]] &=& \left \{ \frac{3\mu_0 \alpha \beta}{4\pi \hat{r}^5}\left[\hat{S}_z^{(p)}(\hat{{\bf S}}^{(l)}\cdot \hat{\bf{r}})+
(\hat{{\bf S}}^{(p)}\cdot \hat{\bf{r}})\hat{S}_z^{(l)} -  \frac{5}{\hat{r}^2}(\hat{{\bf S}}^{(p)}\cdot \hat{\bf{r}})(\hat{{\bf S}}^{(l)}\cdot \hat{\bf{r}})\hat{z}
  + (\hat{{\bf S}}^{(p)} \cdot \hat{{\bf S}}^{(l)}) \hat{z}\right] \right. \nonumber \\
&& \left. +\frac{2\mu_0}{3} \alpha \beta \hat{\textbf{S}}^{(p)} \cdot \hat{\textbf{S}}^{(l)} \frac{\partial }{\partial \hat{z}}\delta^3(\hat{\textbf{r}})\right \} \frac{t^2}{m}~.
\label{commute-2}
\end{eqnarray}

\noindent Note that in (\ref{commute-1}) and (\ref{commute-2}), we have used the general relations

\begin{eqnarray}
[\hat{z}, f(\hat{p}_z)]=i \hbar \frac{d}{d\hat{p}_z}f(\hat{p}_z)
\end{eqnarray}
and
\begin{eqnarray}
[ g(\hat{z}),\hat{p}_z]=i \hbar \frac{d}{d\hat{z}}g(\hat{z})~,
\label{commute-generic}
\end{eqnarray}

\noindent where $f(\hat{p}_z)$ and $g(\hat{z})$ are any functions expandable in a power series of $\hat{p}_z$ and $\hat{z}$, respectively. 

It is evident from (\ref{commute-2}) that $[\hat{O}, [\hat{O}, \hat{z}]]$ is of order $t^2/\hat{r}^4$.  We will show later that the higher order terms in (\ref{zt}) are of order $t^3/\hat{r}^6$, $t^4/\hat{r}^8$, $\ldots$ respectively. To manage this expression perturbatively, we want to be able to ignore these higher order terms. This is possible only if $\left\langle t/ \hat{r}^2 \right\rangle \ll 1$, assuming $\hat{r}$ and $t$ are in units of $l$ and $\tau$, where $l^5={\frac{\mu_0 \alpha \beta \hbar^2}{m}} \tau^2$.  Using $l$ and $\tau$ units is equivalent to replacing $\frac{\mu_0 \alpha \beta \hbar^2}{m}$ with $1$ in our equations.  Note that $t$ is the time interval in which the particle and loop interact.  Since the interaction Hamiltonian is of order $1/\hat{r}^3$, the particle can only have a significant interaction with the loop at small distances from the loop. Therefore, the only way one can ignore the higher order terms is when $t$ approaches zero. In this limit, we can interpret $\left\langle \hat{a}_z\right\rangle$ in Eq.\ (\ref{constant-a}) as a constant acceleration. 


In principle, one may then proceed by computing the expectation values of (\ref{commute-2}) for each of the possible composite spin states, $\Psi_{\uparrow\uparrow}$, $\Psi_{\downarrow\downarrow}$, $\Psi_{\downarrow\uparrow}$, $\Psi_{\uparrow\downarrow}$. The resulting discrete distribution pattern, however, will only distinguish between parallel and anti-parallel dipole configurations. Apparently, this apparatus does not permit a measurement of the individual spins. Hence, we evaluate the expectation values of the mixed states: $\Psi_\parallel=\Psi_{\uparrow\uparrow}+\Psi_{\downarrow\downarrow}$ (if deflected upwards), or $\Psi_{\centernot\parallel}=\Psi_{\uparrow\downarrow}+\Psi_{\downarrow\uparrow}$ (if deflected downwards), see Figure \ref{QubitDetector}. We compute the following expectation values for parallel spins to be
\beeq
{1 \over t^2}\left\langle [\hat{O}, [\hat{O},\hat{z}]]\right\rangle_{\parallel} = \frac{3\mu_0 \alpha \beta \hbar^2}{16\pi m}\left(-5 \left\langle \frac{\hat{z}^3}{\hat{r}^7} \right\rangle + 3 \left\langle \frac{\hat{z}}{\hat{r}^5} \right\rangle \right) +\frac{\mu_0 \alpha \beta \hbar^2}{6m}  \left\langle \frac{\partial}{\partial \hat{z}}\delta^3(\hat{\textbf{r}})\right\rangle~.
\label{z-expect_force}
\eneq
This result is general, and the spatial expectation values can be computed for chosen waveforms of the incident particle. Note, if the spin of the particle is known, we must compute either $\bra{\Psi_{\uparrow\uparrow}}\,[\hat{O}, [\hat{O},\hat{z}]]\,\ket{\Psi_{\uparrow\uparrow}}$ or $\bra{\Psi_{\downarrow\downarrow}}\,[\hat{O}, [\hat{O},\hat{z}]]\,\ket{\Psi_{\downarrow\downarrow}}$, but this gives the same result as Eq.\ (\ref{z-expect_force}). We find the following symmetry holds as one would expect,

\begin{eqnarray}
\left\langle [\hat{O}, [\hat{O},\hat{z}]]\right\rangle_{\parallel} &=& -\left\langle [\hat{O}, [\hat{O},\hat{z}]]\right\rangle_{\centernot\parallel}. \nonumber \\
\end{eqnarray}

In the specific case where $\left\langle \hat{z}^2/\hat{r}^2 \right\rangle=1$, i.e., when the expected positions of the particle and loop are aligned along the $z$-axis, we find
\begin{eqnarray}
{m \over t^2}\left\langle [\hat{O}, [\hat{O},\hat{z}]]\right\rangle_{\parallel}
&=& -\frac{3\mu_0 \alpha \beta \hbar^2}{8\pi} \left\langle \frac{\hat{z}}{\hat{r}^5} \right\rangle~
\end{eqnarray}
after neglecting the term involving the Dirac delta function. This result is consistent with the classical force between two magnetic dipoles, 

\beeq
{\bf F} = - \frac{3\mu_0 m_1 m_2}{2\pi r^5}{z\textbf{k}}~,
\eneq

\noindent where ${\bf m}_1=m_1 \textbf{k}$ (located at the origin ${\bf r}=0$), ${\bf m}_2=m_2 \textbf{k}$ (located on the $z$-axis at $z=r$), and in our case the values of $m_1$ and $m_2$ correspond to $\pm \alpha \hbar/2$ and $\pm \beta \hbar/2$.  

Lastly, we address the commutator terms of order $[\hat{O}, [\hat{O}, [\hat{O}, \hat{z}]]]$ and higher.  Combining (\ref{total-H}), (\ref{bigO}) and (\ref{commute-2}), we find
\begin{eqnarray}
[\hat{O}, [\hat{O}, [\hat{O}, \hat{z}]]] &=& \frac{i}{\hbar} \left [ I \frac{\hat{p}^2}{2m}-\alpha \hat{S}^{(p)}_{z} B_0 -\beta \hat{S}^{(l)}_{z} B_0 - \frac{\mu_0 \alpha \beta}{4\pi \hat{r}^3}\left(\frac{3}{\hat{r}^2} (\hat{{\bf S}}^{(p)}\cdot \hat{\bf{r}})(\hat{{\bf S}}^{(l)}\cdot \hat{\bf{r}})-\hat{{\bf S}}^{(p)} \cdot \hat{{\bf S}}^{(l)}\right) \right. \nonumber \\
&&  \left. -\frac{2\mu_0}{3} \alpha \beta \hat{\textbf{S}}^{(p)} \cdot \hat{\textbf{S}}^{(l)} \delta^3(\hat{\textbf{r}}), \right. \nonumber \\
&& \left. \frac{3\mu_0 \alpha \beta}{4\pi \hat{r}^5} \left(\hat{S}_z^{(p)}(\hat{{\bf S}}^{(l)}\cdot \hat{\bf{r}})+
(\hat{{\bf S}}^{(p)}\cdot \hat{\bf{r}})\hat{S}_z^{(l)} -  \frac{5}{\hat{r}^2}(\hat{{\bf S}}^{(p)}\cdot \hat{\bf{r}})(\hat{{\bf S}}^{(l)}\cdot \hat{\bf{r}})\hat{z}
  + (\hat{{\bf S}}^{(p)} \cdot \hat{{\bf S}}^{(l)}) \hat{z}\right) \right. \nonumber \\
&& \left. +\frac{2\mu_0}{3} \alpha \beta \hat{\textbf{S}}^{(p)} \cdot \hat{\textbf{S}}^{(l)} \frac{\partial }{\partial \hat{z}}\delta^3(\hat{\textbf{r}})\right ] \frac{t^3}{m}~.
\label{commute-3}
\end{eqnarray}
It is straightforward to verify that when commuting $\hat{p}^2=\hat{p}_x^2+\hat{p}_y^2+\hat{p}_z^2$ with the terms of order $1/\hat{r}^4$, one obtains terms of order $1/\hat{r}^6$. Upon commuting terms that involve spin operators in $\hat{O}$, one mostly finds terms of order $1/\hat{r}^7$, which can be neglected compared to the terms of order $1/\hat{r}^6$. The exceptions are $\alpha \hat{S}^{(p)}_{z} B_0$ and $\beta \hat{S}^{(l)}_{z} B_0$, whose spin expectation values contain terms that evaluate to zero. This is expected, since a magnetic field gradient is required to produce a deflection, which the constant $B_0$ terms do not provide. Therefore, it becomes evident that (\ref{commute-3}) is of order $t^3/ \hat{r}^6$ as was claimed earlier.  In the next term, $[\hat{O},[\hat{O}, [\hat{O}, [\hat{O}, \hat{z}]]]]$, it is easy to see that once again we will have $\hat{p}^2$ this time commuting with terms of order $1/\hat{r}^6$, which leads to terms of order $1/\hat{r}^8$.  In other words, our next term will be of order $t^4/\hat{r}^8$.  This trend continues in the susequent terms.

\sxn{Numerical Estimate of the Deflection}
\vskip 0.3cm

\begin{figure}[h]
	\begin{center}
		\includegraphics[height=6cm]{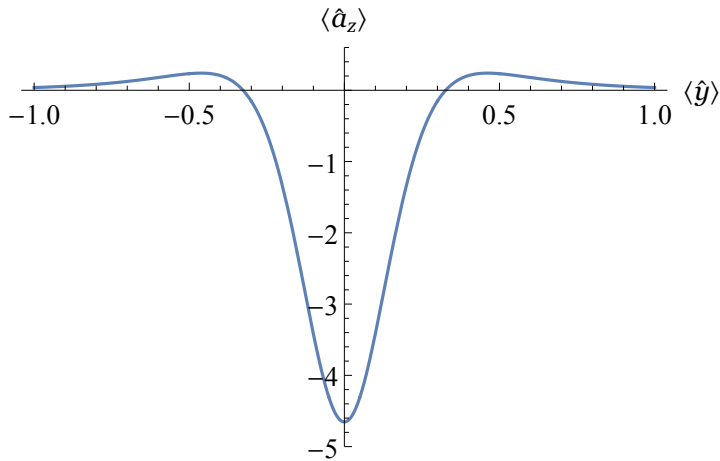}
	\end{center}
	\caption{The expected value of the acceleration in the $z$-direction as a function of $\left\langle \hat{y}\right\rangle$ with $\left\langle \hat{x} \right\rangle = 0$ and $\left\langle \hat{z} \right\rangle = 0.4~l$, where the unit of length is $l=  ({{\mu_0 \alpha \beta \hbar^2} \over {m}}\tau^2)^{1/5}$ and $\tau$ is the unit of time. The horizontal axis is in units of $l$ and the vertical axis is in units of $l/\tau^2$.}
	\label{a-y}
\end{figure}

The general result given in (\ref{z-expect_force}) can be computed for chosen waveforms of the incident particle.  As a numerical approximation, we model the waveform as a square wavepacket of width $0.001~l$, where our unit of length is defined as $l=  ({{\mu_0 \alpha \beta \hbar^2} \over {m}}\tau^2)^{1/5}$ and $\tau$ is the time unit.  We evaluate the expected value of the particle's acceleration, $\left\langle \hat{a}_z\right\rangle$, given in (\ref{z-expect_force}) as a function of $\left\langle \hat{y}\right\rangle$ with $\left\langle \hat{x}\right\rangle=0$ and $\left\langle \hat{z}\right\rangle = 0.4~l$.  The result is shown in Fig.\ref{a-y}.

In the previous section, we indicated that our calculations are valid for infinitesimal time intervals $t$. Nevertheless, one can estimate the deviation of the particle along the $z$-axis by replacing $\left\langle \hat{a}_z\right\rangle$ with its average value found by using the results presented in Fig.\ \ref{a-y}. We find the average acceleration to be $-2.22~l/\tau^2$, where we have ignored the portions of the curve with positive accelerations\footnote{These positive accelerations also occur in classical dipole-dipole interactions at larger distances.}. If the incident particle is a hydrogen atom, and extracted from an oven with a temperature of $100\,^{\circ}$C, one finds its speed to be of order $\sim10^3$ m/s.  The portion of the curve with negative acceleration in Fig.\ \ref{a-y} has an approximate width of $0.6~l$.  Under the assumption that our time unit $\tau$ is in milliseconds, with $m$ as proton mass, $\alpha=-{e\over 2m_e}$, and taking $\beta\sim 10^6 \alpha$, we obtain a unit length of $l\sim 10^{-5}\,$m and consequently find the interaction timescale to be of order $\sim 10^{-8}$ seconds\footnote{To evaluate $\beta$, we assume the current in the superconducting loop is of order $\sim 1~\mu$A and the radius of the loop is of order $\sim 1~\mu$m.}.    This leads to a deviation of order $\sim 10^{-15}$ meters in the negative $z$ direction at the end of the particle-loop interaction, which we assume takes place exclusively in the negative acceleration region.  The anti-parallel dipole configuration will have the same order of magnitude deviation in the positive $z$ direction.   This is a small length and would be experimentally challenging to detect.  However, the conditions might be improved to allow detection using a combination of distant screens, lowering the incident particle's speed, increasing the strength of the magnetic dipole moment of the loop, etc. In addition to this challenge, the separation on the screen needs to be large compared to the width of the wavepacket.  In the case of a hydrogen atom, that would be on the order of an angstrom.   

\sxn{Conclusion}
\vskip 0.3cm

We performed an analysis of the discrete distribution pattern for incoming fermions through a Stern-Gerlach device composed of a two-state magnetic field. The general result can be evaluated for chosen waveforms of the incident particle. We used a square wavepacket to estimate a the order of magnitude of the particle deflection.  The distribution of the particles on the target screen is similar to the classical Stern-Gerlach result. The interesting conclusion with this setup is that if the spin states of the loop and particle remain in a superposition, we can still observe the quantization of spin. In other words, knowledge of the individual spin states is unnecessary; we do not need a classical detector to determine the quantization of intrinsic angular momentum. Our calculations only distinguish between parallel or anti-parallel spin states. 

This does not mean that subsequent measurements of the individual spins are prohibited. One may attempt to measure the spin of the particle directly using an additional mechanism, e.g., having the deflected particle later pass through a classical Stern-Gerlach device. On the other hand, if we assume that the individual spins do collapse upon incidence with the screen, it may be possible to experimentally measure the spin-state of the qubit using a DC superconducting interference device (DC SQUID)\cite{Qubit1, Qubit2}, thereby also learning the spin of the particle. The details of a successful measurement using DC SQUIDS may prove difficult, and are beyond the scope of this project to explore (but see \cite{Qubit3} for a discussion). Nevertheless, this setup may generally prove interesting for investigating the collapse of macroscopic superpositions (see, for e.g., \cite{bassi2013}).

Finally, we find it interesting that one cannot distinguish the individual spin states of the particle or loop by observing the particle distribution on the screen; we only resolve the combined particle and loop state information together. To determine individual spin information of the quantum object measured by our modified Stern-Gerlach device, additional classical detection may be necessary. Moreover, we note that if our proposed macroscopic quantum device was used in Bohm's version of the EPR experiment\cite{BohmEPR} using loops in an {\em{equal}} superposition of mixed magnetic states, the two observers would not establish a correlation between their results (without performing additional classical measurements). For loops in an {\em{unequal}} superposition, however, the correlation may be established. One can obtain the correlation by multiplying the spin state of the EPR particle pair with the spin states of the two loops at the opposite ends of the experiment.  This will give the probability distribution of each spin combination on each screen.  For example, assume the particles of an EPR pair travel in opposite directions, and each couple with a loop in a mixed magnetic state of $10\,\%$ up and $90\,\%$ down. In this case, the $50\,\%$ of the particles that are deflected downward at one end of the experiment will have $82\,\%$ of their pairs deflected upwards at the other end. This suggests that we may be able to establish nonlocality without a classical measurement. 

\sxn{Acknowledgment}
\vskip 0.3cm
We are grateful to Jodin Morey for his contributions to an earlier draft.  We are also grateful to Gianluca Rastelli for informative conversations about qubits and for providing feedback about this paper. 

\def\jnl#1#2#3#4{{#1}{\bf #2} (#3) #4}

\def\Zphys{{\em Z.\ Phys.} }
\def\jssc{{\em J.\ Solid State Chem.\ }}
\def\jpsJ{{\em J.\ Phys.\ Soc.\ Japan }}
\def\ptps{{\em Prog.\ Theoret.\ Phys.\ Suppl.\ }}
\def\PTP{{\em Prog.\ Theoret.\ Phys.\  }}
\def\LNC{{\em Lett.\ Nuovo.\ Cim.\  }}

\def\JMP{{\em J. Math.\ Phys.} }
\def\NPB{{\em Nucl.\ Phys.} B}
\def\NP{{\em Nucl.\ Phys.} }
\def\PLB{{\em Phys.\ Lett.} B}
\def\PL{{\em Phys.\ Lett.} }
\def\PRL{\em Phys.\ Rev.\ Lett. }
\def\PRA{{\em Phys.\ Rev.} A}
\def\PRB{{\em Phys.\ Rev.} B}
\def\PRD{{\em Phys.\ Rev.} D}
\def\PR{{\em Phys.\ Rev.} }
\def\PRe{{\em Phys.\ Rep.} }
\def\PMP{{\em Philos.\ Mod.\ Phys.} }
\def\AP{{\em Ann.\ Phys.\ (N.Y.)} }
\def\RMP{{\em Rev.\ Mod.\ Phys.} }
\def\ZPC{{\em Z.\ Phys.} C}
\def\SCI{\em Science}
\def\CMP{\em Comm.\ Math.\ Phys. }
\def\MPLA{{\em Mod.\ Phys.\ Lett.} A}
\def\IJMPB{{\em Int.\ J.\ Mod.\ Phys.} B}
\def\IJMPA{{\em Int.\ J.\ Mod.\ Phys.} A}
\def\cmp{{\em Com.\ Math.\ Phys.}}
\def\JPA{{\em J.\  Phys.} A}
\def\CQG{\em Class.\ Quant.\ Grav.~}
\def\ATMP{\em Adv.\ Theoret.\ Math.\ Phys.~}
\def\PRSA{{\em Proc.\ Roy.\ Soc.\ Lon.} A }
\def\IJTP{\em Int.\ J.\ Theor.\ Phys.~}
\def\ibid{{\em ibid.} }
\def\LRR{{\em Living \ Rev.\ Relative.} }
\def\FP{{\em Found.\ Phys.} }
\def\BJP{{\em Braz.\ J.\ Phys.} }
\vskip 1cm


\renewenvironment{thebibliography}[1]
        {\begin{list}{[$\,$\arabic{enumi}$\,$]}  
        {\usecounter{enumi}\setlength{\parsep}{0pt}
         \setlength{\itemsep}{0pt}  \renewcommand{\baselinestretch}{1.2}
         \settowidth
        {\labelwidth}{#1 ~ ~}\sloppy}}{\end{list}}

\end{document}